\documentclass[twocolumn]{aastex631}

\usepackage{booktabs,chemformula}
\usepackage{array}

\begin{document}

\title{A Cryogen-Free Electron Beam Ion Trap for Astrophysically Relevant Spectroscopic Studies
\\}

\author[0000-0002-8260-2229]{Amy C. Gall}
\affiliation{ 
 Center for Astrophysics $\vert$ Harvard \& Smithsonian, Cambridge MA
}%

\author[0000-0003-3462-8886]{A. Foster}%
\affiliation{ 
 Center for Astrophysics $\vert$ Harvard \& Smithsonian, Cambridge MA
}%

\author[0000-0002-0167-3245]{Y. Yang}
\affiliation{%
Department of Physics and Astronomy, Clemson University, Clemson SC
}%

\author[0000-0002-2427-5362]{E. Takacs}

\affiliation{%
Department of Physics and Astronomy, Clemson University, Clemson SC
}%
\author[0000-0002-8704-4473]{N. S. Brickhouse}%
\affiliation{  
 Center for Astrophysics $\vert$ Harvard \& Smithsonian, Cambridge MA
}%

\author{E. Silver}%
\affiliation{ 
 Center for Astrophysics $\vert$ Harvard \& Smithsonian, Cambridge MA
}%
\author[0000-0003-4284-4167]{R. K. Smith}%
\affiliation{ 
 Center for Astrophysics $\vert$ Harvard \& Smithsonian, Cambridge MA
}%


\begin{abstract}

The detailed design and operation of the Smithsonian Astrophysical Observatory's EBIT are described for the first time, including recent design upgrades that have led to improved system stability and greater user control, increasing the scope of possible experiments. Measurements of emission from highly charged Ar were taken to determine the spatial distribution of the ion cloud and electron beam. An optical setup consisting of two lenses, a narrow band filter, and a CCD camera was used to image visible light, while an X-ray pinhole and CCD camera were used to image X-rays. Measurements were used to estimate an effective electron density of 1.77 x 10$^{10}$ cm$^{-3}$.  Additionally, observations of X-ray emission from background EBIT gases were measured with a Silicon Lithium detector. Measurements indicate the presence of Ba and Si, which are both easily removed by dumping the trap every 2 s or less. 

\end{abstract}


\section{\label{sec:Intro} Introduction}
Electron beam ion traps (EBITs) are small, tabletop laboratory devices used to create and trap highly charged ions for spectroscopic studies in favorable conditions for astrophysics related investigations. First developed at Lawrence Livermore National Laboratory \citep{Levine_1988}, there are now dozens of EBITs around the world capable of producing astrophysically relevant ions (e.g. \citet{Gillaspy_1995,Silver_1994, Nakamura_1998,Lu_2014, Martinez_2007,Fu_2010}). The physics of these systems has been extensively studied, both experimentally and theoretically, making EBITs a well-understood test bed for benchmarking and testing state-of-the-art atomic structure and collisional population dynamics theories. 

Data from EBITs have been especially important for the accurate interpretation of astrophysical spectra. For example, numerous wavelength measurements of previously unidentified lines have been either added into astrophysical atomic databases used by models or directly used in the interpretation of astrophysical spectra (e.g. \cite{Hell_2016,Brown_1998,Schmidt_2004}). Since the launch of the Atmospheric Imaging Assembly (AIA) on the Solar Dynamics Observatory \citep{Pesnell_2012}, there have been many studies to measure EUV lines missing from the literature that impact the interpretation of measurements from the AIA channels (e.g. \cite{Trabert_2018, trabert_2014b,Trabert_2016}). These results are also relevant for other missions, such as the Extreme Ultraviolet Normal Incidence Spectrograph (EUNIS) sounding rocket \citep{Brosius_2021,Brosius_2014}.
There have also been works to benchmark astrophysically important diagnostic line ratios (\cite{Silver_2000, Chen_2004,Shimizu_2017,Khun_2022}), and to understand the impacts of charge exchange \citep{Wargelin_2008,Shah_2016}, and dielectronic recombination \citep{Bulbul_2019,Gall_2019} within astrophysics.
 \begin{figure*}[ht]
    \centering
    \includegraphics[width = 17 cm, trim={.5cm 1.5cm 1.5cm 1.0cm},clip]{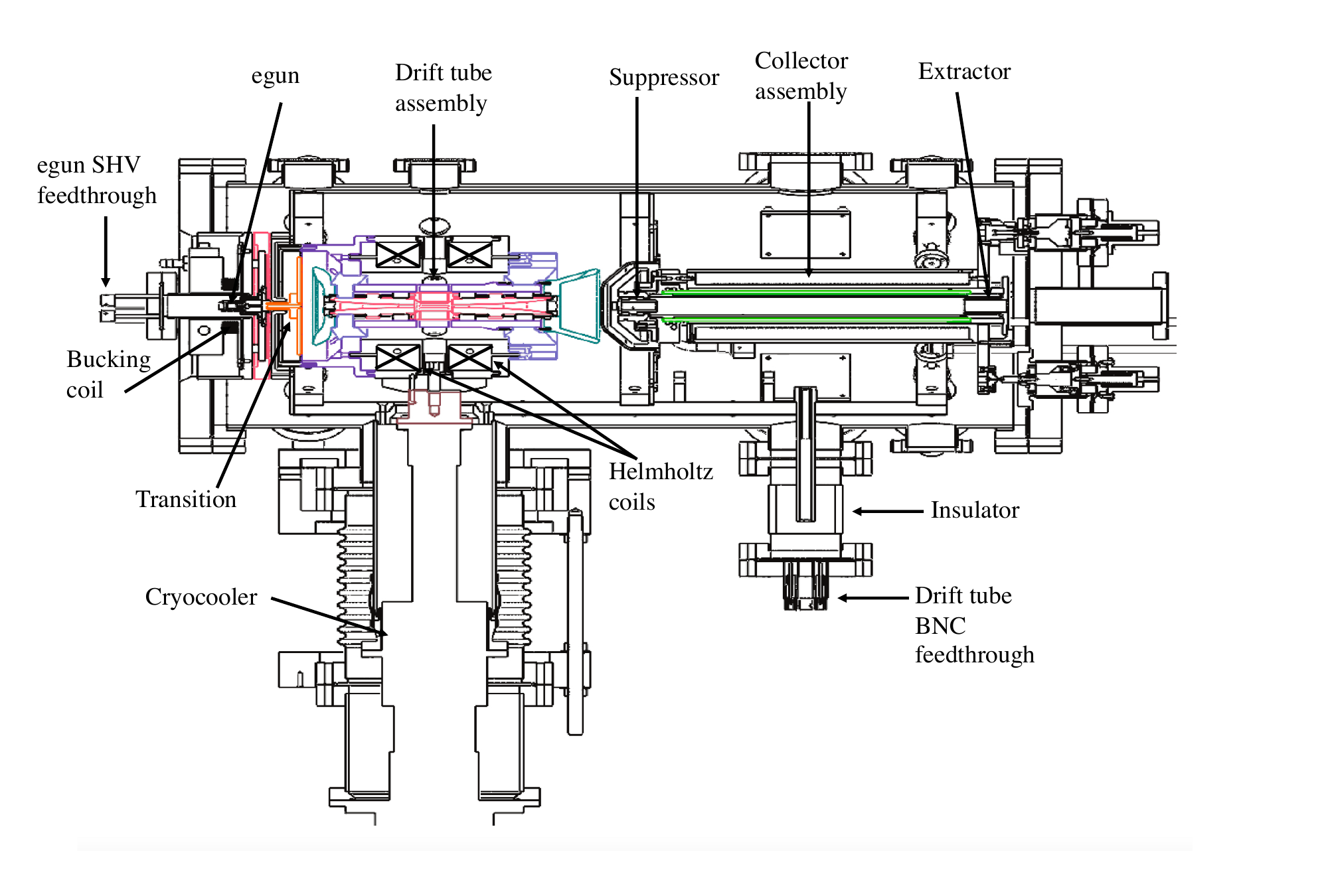}
    
    \caption{ EBIT cross-section view with major components identified.}
    \label{fig:cross-section}
\end{figure*}

The few results from the Hitomi mission \citep{Hitomi_2018}, demonstrated that current X-ray models, such as SAO's AtomDB \citep{Foster_2012,Smith_2001} are not fully prepared for upcoming high-resolution observatories. With the recent launch of the high-resolution X-Ray Imaging and Spectroscopy Mission (XRISM) \citep{XRISM}, new observations will undoubtedly uncover additional deficiencies in the models resulting from incomplete or inaccurate underlying atomic data.  It is critical that laboratories are available to support the data needs of XRISM so scientific objectives can be met. Such laboratories will become even more important as the Athena mission \citep{Barret_2020}, a possible X-ray probe concept, solar EUV missions, such as the Multi-slit Solar Explorer (MUSE) mission \citep{DePontieu2020} and other future missions provide advancements in sensitivity and spectral resolution that push current models beyond their limits. 

Here, we provide the details of the design and operation of the SAO EBIT, located at the Center for Astrophysics $\vert$ Harvard \& Smithsonian, for the first time. Measurements of the electron beam and ion cloud spatial distribution are presented along with a discussion of the effective electron density. Finally, detailed measurements of background elements are presented, which also demonstrate the atomic processes that play an important role in the dynamics of astrophysical and laboratory plasmas.

\section{Background}
The details of the basic design, operation, and physics of EBITs have been described extensively elsewhere \citep{Levine_1988,Currell_2005,Vogel_1990}. In short, a quasi-monoenergetic electron beam emanating from an electron gun ionizes injected and background elements through electron impact ionization. The electron energy distribution typically has a full width at half maximum (FWHM) of about 50 eV, although values as low as 5 eV can be achieved through carefully controlled experimental conditions \citep{Shah_2019}. The tunable beam can be used to probe atomic states and processes such as electron impact excitation, dielectronic recombination, and radiative recombination to name a few. 

The electron beam provides radial trapping of the positive ions and is usually compressed by a superconducting magnet (SCM) to provide high electron densities at the trap. Three trapping electrodes (drift tubes) are used to axially trap ions by placing a higher voltage on the outer electrodes with respect to the middle one. Radial view ports surrounding the trap region allow for spectroscopic observations and injection of neutral elements. 

The horizontally-oriented SAO EBIT was commercially purchased from Physics and Technology Corporation as a laboratory astrophysics facility to support the atomic data needs of current and future high-resolution X-ray missions. One of its main features is that it does not require expensive, and scarce, liquid helium, allowing it to run for long periods of time without interruption. It is a portable and user-friendly device that can be run by a single user. \cite{McDonald_2005} briefly described the device concept and discussed ion extraction applications, something the SAO EBIT is currently not set up to do, but can be potentially added if future need arose. There is one other known EBIT produced from Physics \& Technology. \cite{Schuch_2010} describe design improvements made and additional updates planned for that device. However, the details of the SAO EBIT have not yet been described in any prior publication.  

After its initial commencement, the SAO EBIT has been used at Argonne National Laboratory's Advanced Photon Source to explore photoionization of highly charged Kr ions \citep{Silver_2011}. After returning to SAO, the EBIT experienced significant downtime due to a leak in a collector cooling line that contaminated the vacuum vessel. The instrument has been recently repaired and early experimental tests of the improved system have been described  \citep{Gall_2020_1,Gall_2022,Gall_2022_1}.

\section{\label{sec:BG} Design and Operation of the SAO EBIT}

\subsection{Ion Trap}
\begin{figure}[!ht]
    \centering
    \includegraphics[width = 9.5 cm, trim={1.7cm 2.2cm 2.2cm 2.2cm},clip]{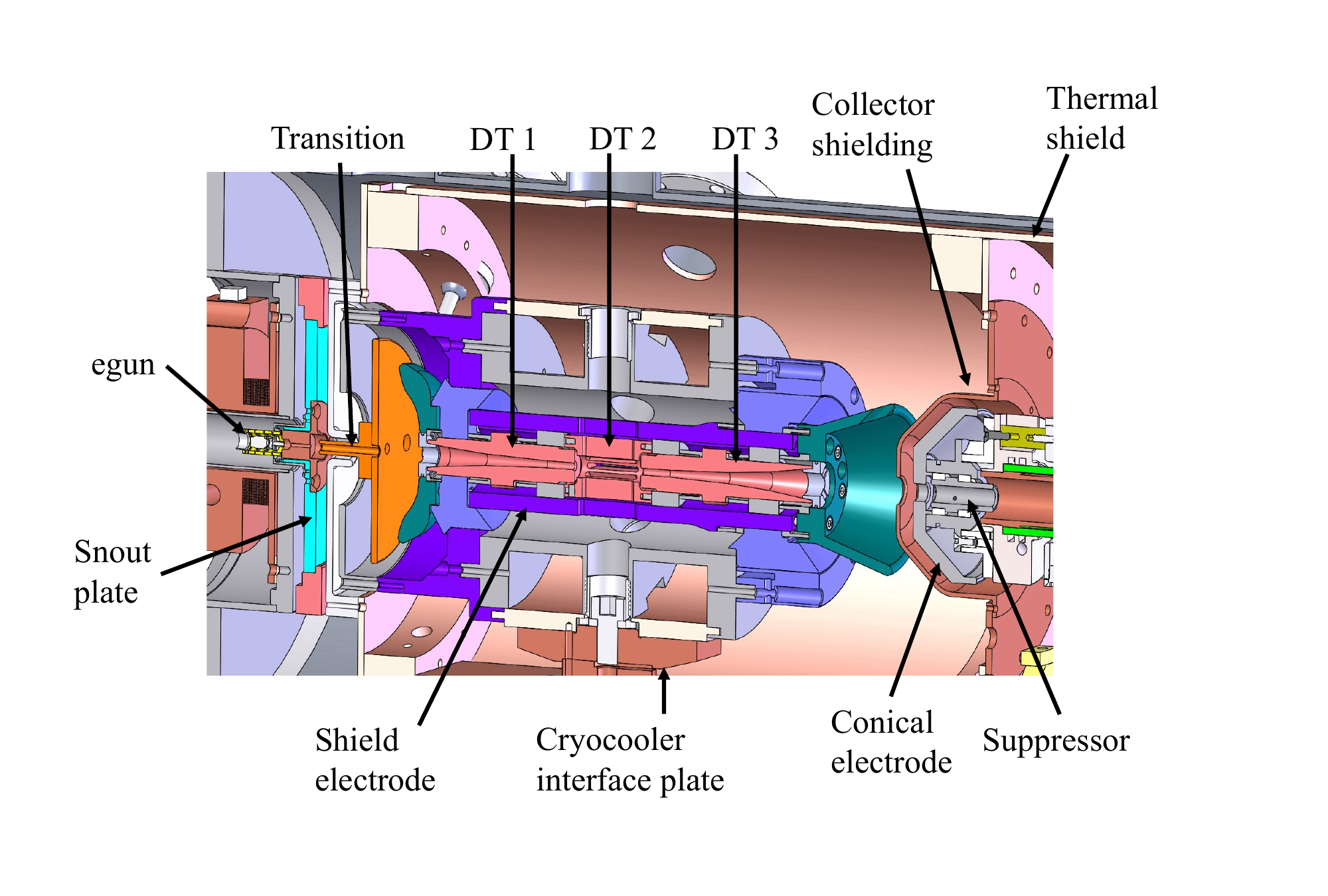}
    
    \caption{Cross-section of the EBIT trapping region. The DTs and shield electrode sit within the SCM housing. Helmholtz coils are not shown, but reside in the two hollow channels shown within the housing. }
    \label{fig:crosszoom}
\end{figure}

The SAO EBIT contains three cylindrical drift tube (DT) electrodes that are used to axially trap ions and accelerate beam electrons (see Fig. \ref{fig:cross-section} and \ref{fig:crosszoom}). The DTs sit within a larger shield electrode and float electrically on top of it. The shield and middle DT contain 8 observation slits that provide access to the trap, while the slit-less outer DTs have a tapered radius to provide an optimum axial potential well \citep{Vogel_1990} (see Fig. \ref{fig:crosszoom}).  

The high voltage (HV) shield electrode power supply has a maximum rating of 40 kV. Recent modifications to the floated drift tube power supply platform allow individual control of each DT. Drift tube 1 (DT1), closest to the electron gun (egun), can float up to 500 V on top of the high voltage of the shield by means of an inexpensive miniature regulated power supply. The middle DT (DT2) and outer DT (DT3) are floated up to 500 V through high voltage amplifiers having a rated slew rate greater than 175 V/$\mu$s for fast switching in dynamical investigations. An 850 nm Luxlink fiber optic system with a 10 ms response time is used to set the voltage of DT1, while faster (5 $\mu$s) systems are used to control DT2 and DT3 amplifiers. The fast amplifiers allow ions to be quickly extracted or dumped from the trap, while the DT1 power supply provides a constant high voltage for trapping and preventing ions from traveling toward the electron gun.

\subsection{Electron Gun}
The EBIT contains a commercially purchased Pierce electron gun with a 3 mm diameter, curved M-type (barium oxide) cathode, a focusing electrode, and an anode. The electron gun can produce up to 150 mA beam currents, and is supported by an Fe snout plate that provides magnetic shielding to the cathode. The anode and snout plate are currently tied together electrically, and share a 5 kV high voltage power supply. Separating the two can potentially allow for fast manipulation of the electron beam for atomic lifetime measurements or other dynamical studies. 

A bucking coil surrounds the electron gun and optimizes the magnetic field at the cathode to reduce the beam radius \citep{Takacs_1996}. The coil is outside of the vacuum chamber and cooled with an inhibited technical-grade ethylene glycol. The cathode rests at ground potential, so the nominal electron beam energy at the trap (not corrected for the space charge of the electrons and the trapped ions) is determined by the voltage of the DT2, with respect to earth ground.

\subsection{Collector}
The collector assembly was redesigned following a failure in a brazed joint connecting the cooling line to the collector body. The silicone oil coolant used at the time contaminated the vacuum chamber, requiring total disassembly, cleaning, and replacement of damaged components. The new design includes welded joints, the addition of a suppressor electrode which was previously absent, and improved electrical isolation between electrodes. 

A flat-top, conical electrode at the end of the assembly is electrically isolated from the suppressor and collector, but is currently tied to the suppressor, sharing a 5 kV power supply. The copper collector electrode has a 2 kV power supply and is cooled with a non-conductive, residue-free coolant that evaporates in air and will not expand if frozen during a power outage. A coil surrounding the collector provides a magnetic field that spreads out the dense electron beam for collection with efficient power dissipation.

\subsection{Control System}
A new high speed analog output (AO) device is used to control the EBIT electrode voltages, while an analog input device is used to monitor voltages and currents. The 16-bit, 32 channel, 1 MS/s AO device is capable of periodic and non-periodic waveform output. A user interface was created in Python programming language to set static voltages and to monitor electrode voltages and currents in real time. It is also used to program voltage ramps for dumping and injecting ions, and quickly switching between different beam energies. As a safety feature, the program continuously monitors the anode current and voltage, sets the voltage to zero if a user-defined power threshold is met, and sends an alert to the user. The AO device is also capable of producing Transistor Transistor Logic (TTL) triggers to synchronize EBIT operations with external devices and detectors. 

\subsection{Superconducting Magnet and Thermal System}
The superconducting magnet (SCM) consists of a conduction cooled assembly containing a pair of Helmholtz coils and eight 0.7" (17.78 mm) radial access ports. The SCM is capable of producing a 3 T magnetic field at 62 A current and has $\pm$ 0.1\% homogeneity over a 1 cm diameter spherical volume. The SCM is cooled to a temperature of less than 4K by the second stage of a Sumitomo Heavy Industries (SHI) cryocooler. The cryocooler has a refrigeration power of 50 W at 40 K temperature (first stage) and 1 W power at 4.2 K temperature (second stage). 

The thermal shield separating the room temperature vacuum chamber from the SCM is cooled below a temperature of 45 K by the first stage of the SHI cryocooler and by a second, single state Cryodyne cryocooler (80 W at 77 K capacity) that was later added for additional cooling. The high temperature superconducting current leads of the SCM attach to copper wires that travel through a vacuum feed-through flange to the power supply. The copper wires contact the 45 K thermal shield through a thermally conductive polyimide film to create a smooth temperature gradient between the room temperature feed-through and the 4K SCM. 

A new interface plate was installed between the SCM and cryocooler that more than doubled the contact surface area. A thin gold film was placed between the cold head plate and SCM interface to further maximize contact. To reduce radiative heating from the warm collector, a flat-top conical copper shield, attached to the 45 K EBIT thermal shield, was placed between the SCM and the collector assembly (see Fig. \ref{fig:crosszoom}).

\begin{figure}[!ht]
    \centering
    \includegraphics[width = 8 cm]{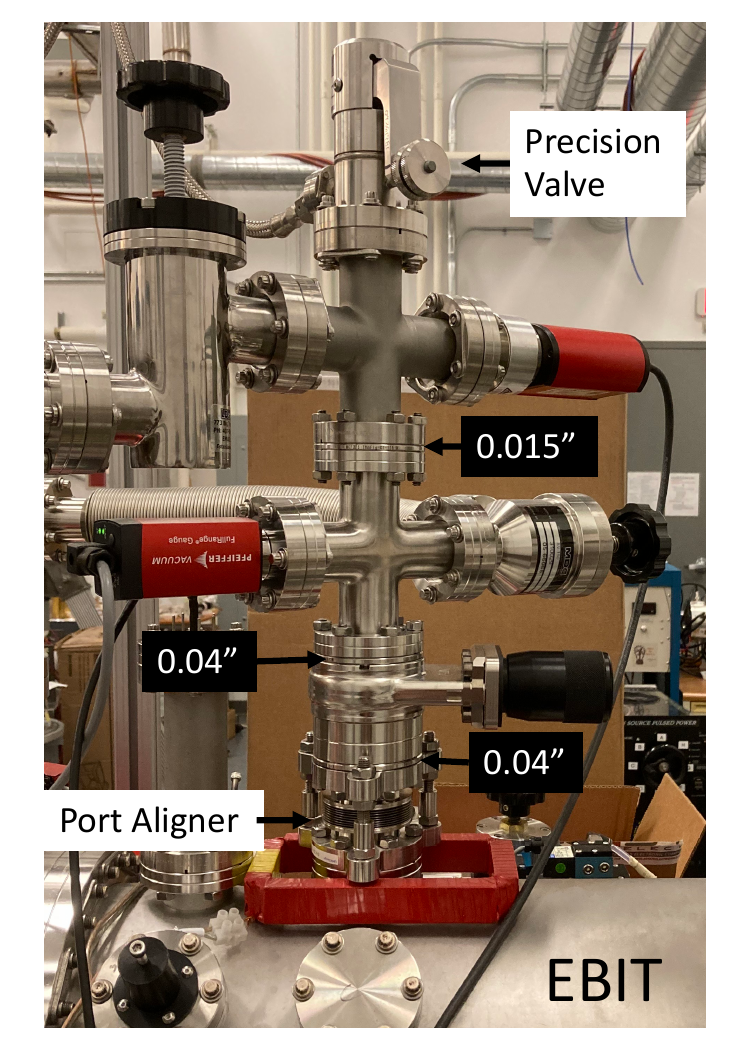}
    \caption{Differentially-pumped ballistic gas injection system placed on the top EBIT observation port. Black text boxes indicate aperture radii. Red coil at the EBIT port is one of three steering coils surrounding the trap.}
    \label{fig:gas}
\end{figure}

\subsection{Element Injection}
A differentially-pumped ballistic gas injection system was recently added to the EBIT to introduce a well-collimated beam of gas into the trap (see Fig. \ref{fig:gas}). A gas manifold containing 4 selectable gases and a pump-out line is connected to a turbomolecular pumping station. Gas of neutral atoms of choice is introduced into a 2.75" (69.85 mm) conflat flange 4-way cross through an adjustable rate, precision valve. The 4-way cross is pumped with a turbo station (67 l/s N$_{2}$ pumping speed) through an inline valve. Gas flows down through a 0.015" (0.381 mm) radius aperture into a second 4-way cross, pumped with an identical turbo station through a right angle valve. The gas then flows through two additional apertures with 0.04" (1.02 mm) radii into the EBIT trapping region. Pressure in the top section is typically 100 times higher than the pressure in the lower section. A port aligner setup is used to align the gas jet with the drift tube slits. A gate valve separates the gas injection system from the EBIT vacuum when not in use. 

A Metal Vapor Vacuum Arc (MeVVA) system, almost identical to the system built for the NIST EBIT, was purchased from the Naval Research Laboratory (NRL) \citep{Holland_2005}. Fe, Ti, Ni, Mg, and Zn cathodes are currently installed, but any element that can be formed into a stable solid cathode can be installed. Ions created by the MeVVA's floated anode and cathode travel toward the EBIT trap along magnetic field lines. During injection, the EBIT's shield high voltage matches the MeVVA voltage, so that ions will have zero kinetic energy once they reach the trap for confinement.

\subsection{Operation of the EBIT}

The EBIT was aligned at room temperature by viewing crossed strings, placed on the two ends of the EBIT vacuum chamber and on the ends of the horizontal drift tubes, through an alignment telescope. The bottom of the SCM, and trapping electrodes within the SCM, sit directly on the cryocooler coldhead. Therefore, alignment of the drift tubes is accomplished by mechanically moving the coldhead up and down on three, equally spaced, threaded rods anchored to the EBIT flange. Once aligned, the drift tubes were raised vertically an additional 0.025" (0.635 mm) to account for material contraction when cold. Cooling the EBIT to the operating temperature is accomplished with the two cryocoolers and typically requires less than 24 hours. 

During operation, the anode, transition, high voltage, and collector current readouts are used to tune the electron beam. Electrode voltages, the collector coil, and the bucking coil are tuned to maintain an anode power below 0.5 W, although the anode current can usually be tuned close to zero at the expense of a lower emission current. 

\begin{figure}[!ht]
    \centering
    \includegraphics[width = 9.0 cm]{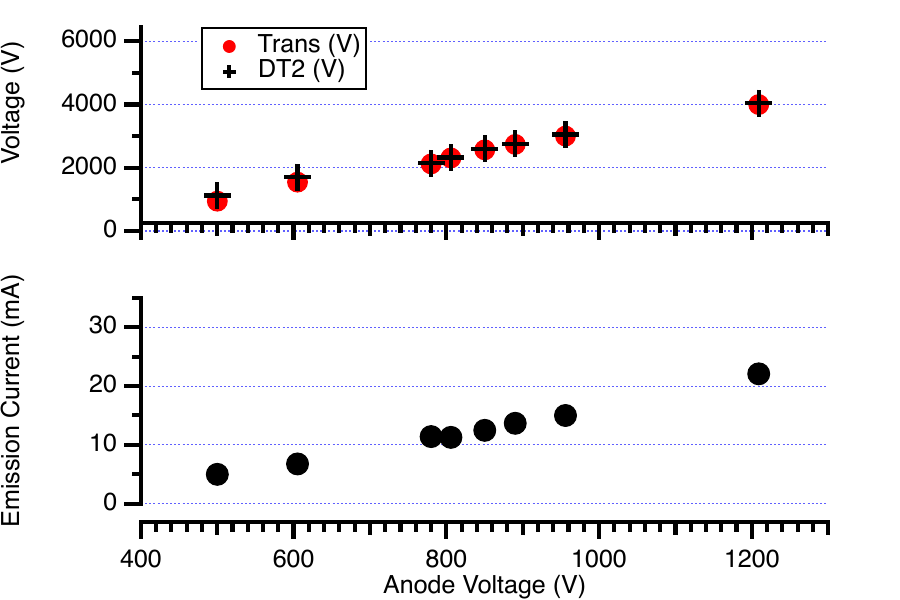}
    \caption{Transition, anode, and electron beam energy settings used for the measurements described in Section \ref{sec:BGmeas}. Top.) Transition voltage required to produce a near-zero anode current at each anode voltage setting (red points). The DT2 settings are also shown at each point and are nearly equal to the optimum transition setting.  Bottom.) Emission current resulting from each anode voltage setting. Higher currents can be achieved by relaxing the 1$\mu$A anode current limitation.}
    \label{fig:beamcurrent}
\end{figure}

The bucking and collector coils are tuned to minimize the difference between the emission and collector currents and to optimize the electron density for efficient ion production in the trap region. The physical position (alignment) of the collector and drift tubes can also be adjusted, as needed, until the emission and collector currents are best matched. Steering coils surrounding the trap are also available for tuning, but are not typically needed.

As mentioned, DT2 floats on top of the shield electrode HV and determines the electron beam energy. The anode voltage setting determines the electron emission current, and is typically optimized at values less than the DT2 voltage setting (see Fig. \ref{fig:beamcurrent}). The beam is generally more stable at higher electron beam energies ($>$ 2 keV), while at lower energies, a reduced magnetic field and lower beam current are used to reduce the excess anode and transition currents. 

During the measurements described in Section \ref{sec:BGmeas}, the transition voltage was tuned until the anode current was minimized to less than 1 $\mu$A. Over the DT2 voltage range of 1130 - 3050 V, the optimum transition voltage was typically equal to the DT2 voltage (see Fig. \ref{fig:beamcurrent} (Top)). The EBIT electrodes can be tuned to produce higher emission currents than those shown in Fig. \ref{fig:beamcurrent}, at the expense of higher anode currents. At higher, more stable electron beam energies, a higher magnetic field can be used and the optimum transition voltage (limited to 5 kV) is no longer equal to the electron beam energy.

Background trace contaminants, described in Section \ref{sec:BGmeas}, including Ba from the electron gun, are present within the EBIT and become ionized and trapped. Tests performed by injecting Ar at 5.4 $\times$ $10^{-7}$ Torr (measured in the lower section of the gas injection system) showed that by emptying the trap, every 2 seconds or less, signatures from contaminants are removed from the spectrum without degrading the injected Ar signal.\\
\clearpage
\section{\label{sec:imaging} Electron beam and ion cloud imaging}

Electron densities in solar and astrophysical sources can be determined spectroscopically through the ratio of two lines, one (or both) of which is connected to a metastable level \citep{Del_zanna_2018}. The lines are usually from the same ion and close in wavelength to avoid uncertainties due to ion populations and different temperature dependencies. Since the models used to interpret spectra rely on theoretical atomic data, it is important to verify these diagnostic ratios in a well understood laboratory environment.

Typical electron densities found within an EBIT are comparable to those of the Solar Corona, making them ideal devices for benchmarking density diagnostics that fall within the bandpass of observatories such as the Extreme Ultraviolet Imaging Spectrometer aboard Hinode  (see studies by e.g., \cite{Shimizu_2017, Nakamura_2021,Nakamura_2011,Arthanayaka_2020}). 
To carry out density calibration studies, it is crucial to have an accurate understanding of the EBIT electron density. The density can be calculated based on the assumption of a uniform cylindrical beam as:
\begin{equation}
n_{{\rm e}}=\frac{I_{{\rm e}}}{ev_{{\rm e}} \pi r_{{\rm e}}^2}
\label{eq:0}
\end{equation}

Here $e$ is the charge of an electron and $I_{{\rm e}}$, $v_{{ \rm e}}$, and $r_{{\rm e}}$  are the electron beam current, electron velocity, and electron beam radius, respectively. 
However, the ion cloud is typically much larger than the electron beam so ions 
can experience a lower effective electron density as their trajectories take them outside of the beam. 
Moreover, the electron beam and the ion cloud are found to be better approximated by Gaussian distributions (e.g. \cite{Knapp_1993,Nakamura_2021}). Therefore, an effective electron density, that considers the distributions of both the ion cloud and electron beam, is represented as \citep{Shimizu_2017}:

\begin{equation}
n_{{\rm e,eff}}=\frac{4\ln(2)I_{{\rm e}}}{\pi(\Gamma_{{\rm e}}^2+\Gamma_{{\rm i}}^2)ev_{{\rm e}}}
\label{eq:1}
\end{equation}

Where $\Gamma_{{\rm e}}$ and $\Gamma_{{\rm i}}$ are  the FWHM of the electron beam and ion cloud. 

The spatial distributions of the electron beam and the ion cloud are highly dependent on experimental conditions. For example, \cite{Porto_2000} showed that the size of the ion cloud increases with trap depth, resulting from the increased temperature required to leave the trap, while \cite{Liang_2009} found that the ion cloud size depends on the charge state, with lower charge states being more expanded. \cite{UTTER_1999} demonstrated that the measured electron beam radius changes with electron beam current and with the magnetic field produced by the bucking coil and superconducting magnet. These works suggest that density dependent investigations require measurement of the spatial distribution of the ion cloud and electron beam simultaneously with density sensitive lines, such that the effective electron density can be accurately determined. 
\begin{figure*}[ht]
  \centering
  \includegraphics[width=0.45\textwidth]{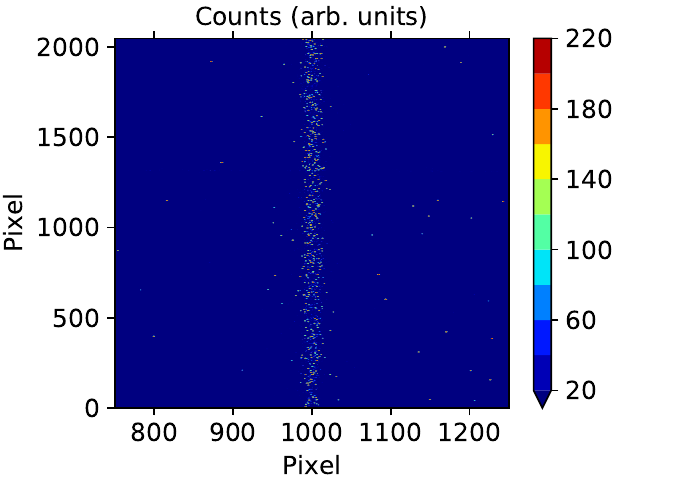}\label{fig:xrayimg}
  \hfill
  \includegraphics[width=0.45\textwidth]{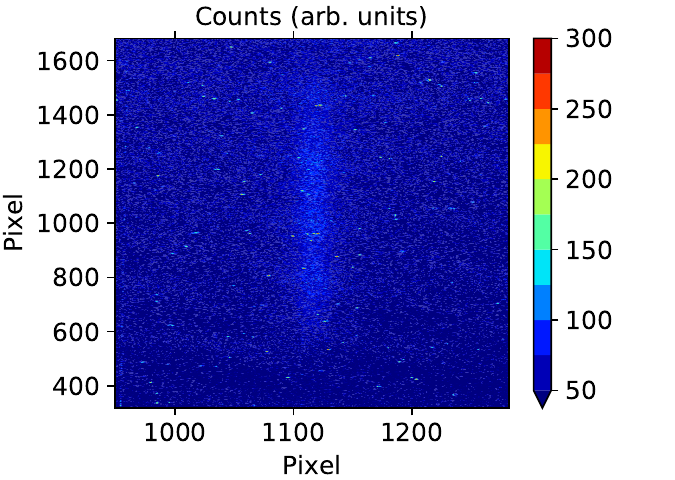}\label{fig:visimg}
  \caption{(Left) Measured X-ray emission from highly charged Ar representing the spatial distribution of the electron beam. (Right) Measured optical, 441 nm light from Ar$^{13+}$ representing the spatial distribution of the ion cloud.}
  \label{images}
\end{figure*}

To this end we designed a setup to measure the spatial distributions of the electron beam and the ion cloud at the SAO EBIT to characterize the system and understand typical effective electron densities that can be achieved in laboratory astrophysics studies.

\subsection{\label{sec:vis}Visible imaging details }
The boron-like $2s^22p$\ 
$^2P_{1/2}$ $-$ $^2P_{3/2}$ forbidden magnetic dipole (M1)  transition in Ar$^{13+}$ ions \citep{Orts_2006} at 441.2 nm allowed the imaging of the ion cloud.
Using $kT_i \approx 0.3q V_{ax}$ to estimate the ion temperature from the axial potential ($V_{ax}$), defined by the voltage setting of DT3, and ion charge (q) \citep{Currell_2003}, we find the ion radial motion occurs on the order of 1 ns (assuming a 100 $\mu$m beam radius). The lifetime of the Ar$^{13+}$ transition is on the order of 10 ms \citep{Serpa_1998,Lapierre_2006}. Therefore, the ions can sample the entire ion cloud, and the spatial distribution of the measured emission represents the spatial extent of the ions.   

Visible measurements were taken using two coaxial achromatic doublet lenses, a 440 nm filter with a 10 nm FWHM bandwidth, and an Eagle 42-40 CCD camera. A 2.5 mm thick sapphire window mounted on a conflat flange separated the EBIT vacuum from the optical enclosure at atmosphere. The EBIT source was placed at the focus of a 500 mm focal length lens while the CCD imagining plane was placed at the focus of a 400 mm focal length lens. The filter was placed between the two lenses where the light was collimated. The system was aligned by shining a laser through the EBIT, perpendicular to the electron beam direction, through the center of two coaxial observation ports, and through the center of the EBIT trap. The position of the lenses, filter, and CCD were adjusted until the light hit the center of each element.

The lenses, filter, and CCD were mounted on an optical breadboard using standard optical hardware. An optical enclosure box was constructed with black construction rails mounted on the breadboard. Laser-cut, black coated stainless-steel plates were used to attach the CCD and EBIT conflat flange to the box. The remaining sides of the enclosure were made from thick black posterboard. As an additional precaution to prevent environmental stray light from entering the enclosure, measurements were taken in a darkened laboratory.

The optical configuration resulted in a magnification of 0.80, verified by imagining the drift tube slit without the filter in place. With a 13.5 $\times$ 13.5 $\mu$m pixel size, this magnification resulted in an object size of 16.875 ${\rm\mu}$m per pixel.

Visible measurements were taken at a nominal electron beam energy of 6050 eV and beam current of 33 mA. The estimated shift in energy due to the space charge is about 15 eV \citep{Porto_2000}, resulting in less than a 0.5\% change in the final effective density, and so is ignored in this case. The superconducting magnet was set to 1 T while the bucking coil was tuned to minimize current on the anode while maximizing the current received at the collector. Argon was injected using the gas injection system at a pressure of 1.6 $\times$ $10^{-6}$ hPa. The trapping electrodes were set to 400 V, 0 V, and 150 V for DT1, DT2, and DT3, respectively. To remove background contaminates, described in Sec. \ref{sec:BGmeas}, the trap was dumped every 2 s by increasing the DT2 to 250 V for a 5 ms resulting in the ejection of the ions in the direction of DT3. Four frames were collected in this manner with an exposure time of 30 minutes each. 

Background measurements, taken with the electron gun off, were subtracted from each frame to remove any residual light from the filament and a fringe pattern caused by the sapphire window. The frames were summed and rotated by 0.25\textdegree\ to correct for misalignment of the CCD pixel array and the object plane. CCD pixels that accumulated very high electron counts were eliminated from further analysis as these corresponded to cosmic rays striking the detector during the image acquisition. Pixel columns, along the straight image of the ion cloud, were summed together to analyze the radial ion cloud profile.

\begin{figure*}[ht]
  \centering
  \includegraphics[width=0.45\textwidth]{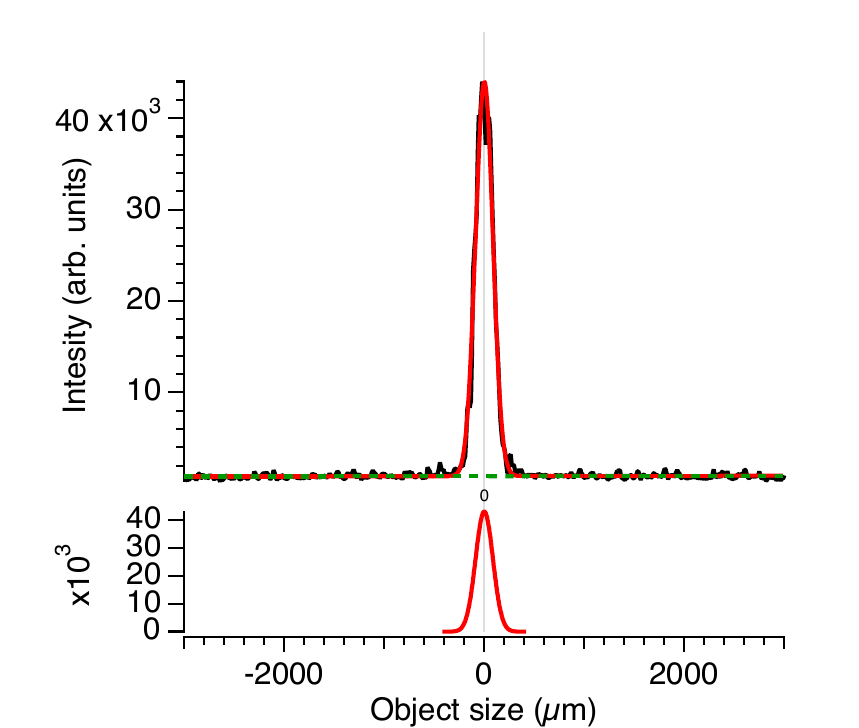}
  \hfill
  \includegraphics[width=0.45\textwidth]{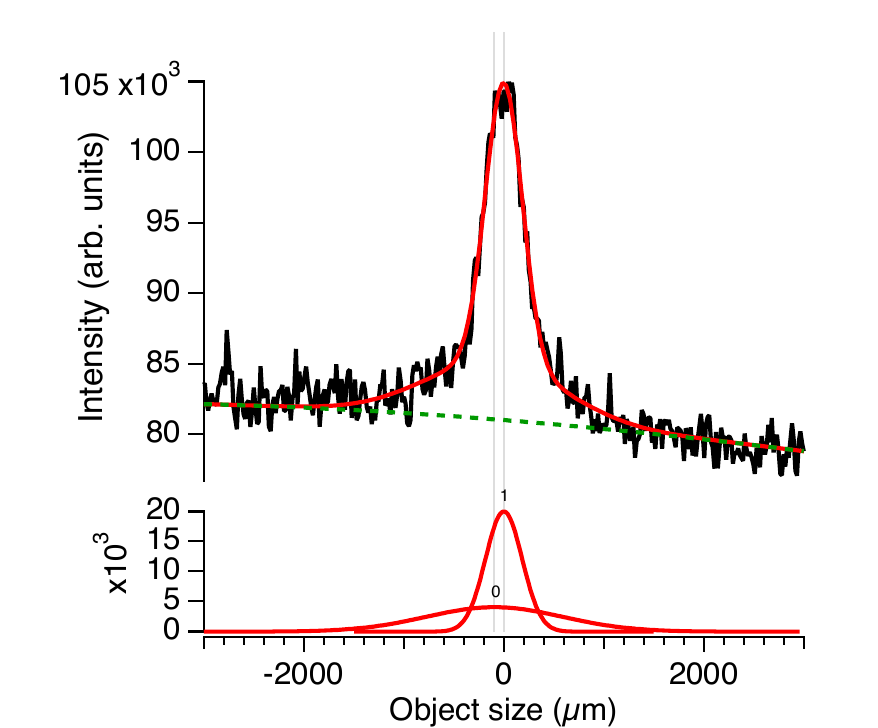}
  \caption{Left) Collapsed X-ray image. Right) Collapsed visible image. }
  \label{fig:collapsed}
\end{figure*}

\subsection{\label{sec:x-ray}X-ray imaging measurements}
The electron beam was imaged by detecting X-ray emission from highly charged, primarily Ar$^{16+}$ and Ar$^{17+}$, argon ions. With typical lifetimes on the order of 10$^{-10}$ s, X-ray emission from these ions occur within the electron beam, providing a measure of the spatial distribution of the electrons in the trap region. A 20 mm long, 50 $\mu$m width slit was used to image the X-ray emission. The laser cut slit plate was made from 0.1" (2.54 mm) thick, 304 stainless steel and mounted on a custom 0.175" (4.445 mm) thick conflat flange. The slit was positioned 420.7 mm from the EBIT source and 498.2 mm from the imagining plane, resulting in a geometric magnification of 1.18. The slit assembly was placed under vacuum and separated from the EBIT through a 200 nm thick polyimide film. A 2048 $\times$ 2048 pixel, Andor Ikon-L CCD camera was used to image the electron beam. The 13.5 $\times$ 13.5 $\mu$m pixel size results in an object size of 11.44 $\mu$m per pixel.  

Measurements were made using the same experimental conditions described in  Sec. \ref{sec:vis}. A single 15 minute X-ray emission exposure was followed by a 15 minute background measurement with the electron gun turned off. The background subtracted image was rotated 0.33\textdegree\ to correct for slight misalignment of the slit and the imaging plane. Cosmic rays were removed using energy and spatial discrimination techniques described by \cite{HUDSON_2007}. Pixels were summed along the length of the X-ray image to determine the size of the beam.

\subsection{Spatial distribution analysis and results}

The processed visible and X-ray images are given in Fig. \ref{images}, where the pixels have been scaled such that both images have roughly the same magnification for comparison. Emission from $n=2-1$ transitions in H-like and He-like Ar ions are expected to dominate the X-ray image, with an energy of $\sim$ 3 keV. The collapsed X-ray image profile, shown in Fig. \ref{fig:collapsed} (Left), was fit using a single Gaussian and a constant background, resulting in a measured FWHM of 204.8 $\pm$ 1.07 $\mu$m, where uncertainties are at a 1$\sigma$ confidence level. The size of the X-ray image includes contributions from the finite size of the slit. With emission primarily from photons with energies $>$3 keV, the contribution from diffraction is small ($<$ 1\%) and ignored in this case. The size of the beam is then defined as:

\begin{equation}
\sigma_{\text{beam}} = \sqrt{\sigma_{\text{beam(meas)}}^{2}-\sigma_{\text{slit}}^{2}}
\label{eq:2}
\end{equation}

Where $\sigma_{\text{beam}}$ is the true electron beam size, $\sigma$$_{\text{beam (meas)}}$ is the measured image size, and $\sigma$$_{\text{slit}}$ is the contribution from the slit. This contribution is estimated from the standard deviation of a uniform distribution as $\sigma_{\text{slit}} =\frac{w (d_1+d_2)}{\sqrt{12}d_1}$, where $d_1$ is the distance between the slit and EBIT source, $d_2$ is the distance between the slit and image plane, and $w$ is the width of the slit \citep{TAKABAYASH_2013,Elleaume_1995}. 
Using a similar setup, \cite{Yang_2011} used large scale ray tracings to estimate the broadening effects caused by the finite slit width. Comparison of their measured and corrected beam width is consistent with using $\sigma_{\text{slit}}= \frac{w (d_1+d_2)}{\sqrt{12}d_1}$ to within less than 8.6\%, providing a rough estimate of its uncertainty.

The estimated corrected electron beam size (FWHM) is $\Gamma_\text{e}$ = 190.87 $\pm$ 17.16 $\mu$m. Measurements were also taken with the same experimental conditions, without injection of Ar and without dumping the trap. Emission from background elements (described in Sec. \ref{sec:BGmeas}) formed the image of the electron beam and resulted in a final FWHM of 203.0 $\pm$ 17.98 $\mu$m, within the uncertainty of the Ar measurement.  While these measurements provide a rough estimate of the beam size, followup studies varying the slit width can be performed to accurately determine the contribution from the slit (e.g. \cite{Baumann_2014}). Alternatively, a high-resolution, slit-less spectrometer can provide a measurement of the electron beam without the need of a slit or pinhole (e.g. \cite{Arthanayaka_2018}).

\begin{table*}[t]
\centering
\caption{Summary of imaging measurements including key experimental parameters, FWHM of the ion cloud and electron beam, and densities. }
\begin{center}
\begin{tabular}{cccccccc}
\toprule
$E_\text{e}$ {[}eV{]} & $I_\text{e}$ {[}mA{]} & B-field {[}T{]} & $\Gamma_\text{e}$ {[}$\mu$m{]} & $\Gamma_{\text{ion,n}}${[}$\mu$m{]} & $\Gamma_{\text{ion,w}}$ {[}$\mu$m{]} & $n_{\text{e,eff}}$ {[}$\text{cm}^{-3}${]} & $n_{\text{e}}$ {[}$\text{cm}^{-3}${]} \\
\midrule
6050        & 33          & 1.0             & 190.87 $\pm$ 17.16        & 431.73 $\pm$ 11.86         & 1602.7 $\pm$ 119.29           & 1.77 x 10$^{10}$          & 1.5 x 10$^{11}$      
\end{tabular}
\label{tab:results}
\end{center}
\end{table*}

Fitting the visible peak from the collapsed image with a single Gaussian and cubic background resulted in a FWHM of 556.95 $\pm$ 8.6195 $\mu$m. However, the fit at the base of the peak was poor. A better fit resulted from a broad Gaussian base and a second narrower Gaussian peak as shown in Fig. \ref{fig:collapsed} (Right). The FWHM of the wide ($\Gamma_{{\rm ion,w}}$) and narrow ($\Gamma_{{\rm ion,n}}$) Gaussian peaks are 1602.70 $\pm$ 119.29 and 431.73 $\pm$ 11.86 $\mu$m, respectively. 

This shape has also been observed in other EBIT measurements of the ion cloud (e.g. \cite{Porto_2000,Shimizu_2017, Arthanayaka_2018, Nakamura_2021}). \cite{Porto_2000} modeled a thermal distribution of ions within an electrostatic potential. Their truncated Boltzmann distribution showed reasonable agreement with measurements.  \cite{Arthanayaka_2018} attributed the narrow peak to target emission, and the broad peak to aberrations in the lens and misalignment of the lens and object plane. They independently verified this by imaging a slit while adjusting the lens plane and found the narrow slit to be consistent while the centroid and width of the broad peak shifted. In our fit, the broad peak is shifted $\sim$ 90 $\mu$m from the narrow peak, demonstrating a similar behaviour.

Using the narrow Ar peak as the ion cloud FWHM ($\Gamma_{{\rm ion,n}}$), the effective density was found to be 1.77 x 10$^{10}$ cm$^{-3}$, roughly an order of magnitude less than predicted from a homogeneous cylindrical beam (see Table \ref{tab:results}).

The 441 nm, Ar$^{13+}$ transition has been used to probe the ion cloud size at other EBIT facilities as well. \cite{Nakamura_2021} took measurements with two EBITs with different experimental conditions and measured the image profile with two Gaussians, $\Gamma_1$ = 130 $\mu$m and $\Gamma_2$ = 490 $\mu$m, with the Tokyo-EBIT, while a single Gaussian, $\Gamma$ = 690 $\mu$m, fit best to the measured profile from the CoBIT. \cite{Porto_2000} measured the cloud width as $\sim$ 300 $\mu$m at the same trap depth used in this work. The range of values demonstrates the sensitivity of the ion cloud size to experimental parameters, such as the beam current, the magnetic field at the cathode and trap, and the presence of light species in the trap region that evaporatively  cool heavier ions. These can vary greatly from one experiment to another and can provide flexibility over the ion cloud conditions to study.

\section{\label{sec:BGmeas}Measurements of emission from trace background elements}
\subsection{\label{sec:setup} Experimental Setup}
To identify background ions present in the EBIT, X-ray observations of the EBIT plasma were made with a lithium-drifted silicon (Si(Li)) detector. The broadband X-ray detector has an active area of 10 mm$^2$ and an energy resolution of about 150 eV at 6 keV. Measured spectra were energy calibrated using an external X-ray fluorescence calibration source that produces low-background, characteristic X-rays by exposing target materials to bremsstrahlung radiation emanating from an X-ray tube with a tungsten filament and copper anode. Low-energy photons from the X-ray tube are blocked by a 50 nm Al and 135 nm polyimide filter. The calibration source and Si(Li) detector were placed under vacuum and separated from the EBIT through a 200 nm polyimide filter.

Measurements were taken in a steady state mode, where the electron beam energy was kept fixed and incrementally increased from 1285 eV to 3095 eV in steps of 15 eV, with collection times between 15 - 30 minutes at each beam energy setting. A magnetic field of 0.5 T was set to maximize the stability of the electron beam at the lowest energy and used throughout.  

The focus, suppressor, extractor, and collector electrodes were set to -4 V, 800 V, -500 V, and 1200 V, respectively, to minimize the anode current and maximize the collector current. The current measured on the collector ranged from 4.9 mA at the lowest beam energy to 14.7 mA at the highest (see Fig. \ref{fig:beamcurrent}). The signal from dielectronic resonances (DR) was increased by periodically emptying the ions from the trap as they strongly affect the ion charge distribution. While dumping, the voltage of DT2 was raised above DT3 for 5 ms at every 60 s, as described in Sec. \ref{sec:dump}. No gas or element injection was used, so emission was solely due to lingering background gas and EBIT contaminants, including a small amount of Si residue after thorough cleaning the leaked collector coolant and Ba emanating from the BaO${_2}$ coated electron gun cathode.

\begin{figure*}[ht]
    \centering
    \includegraphics[width=17cm]{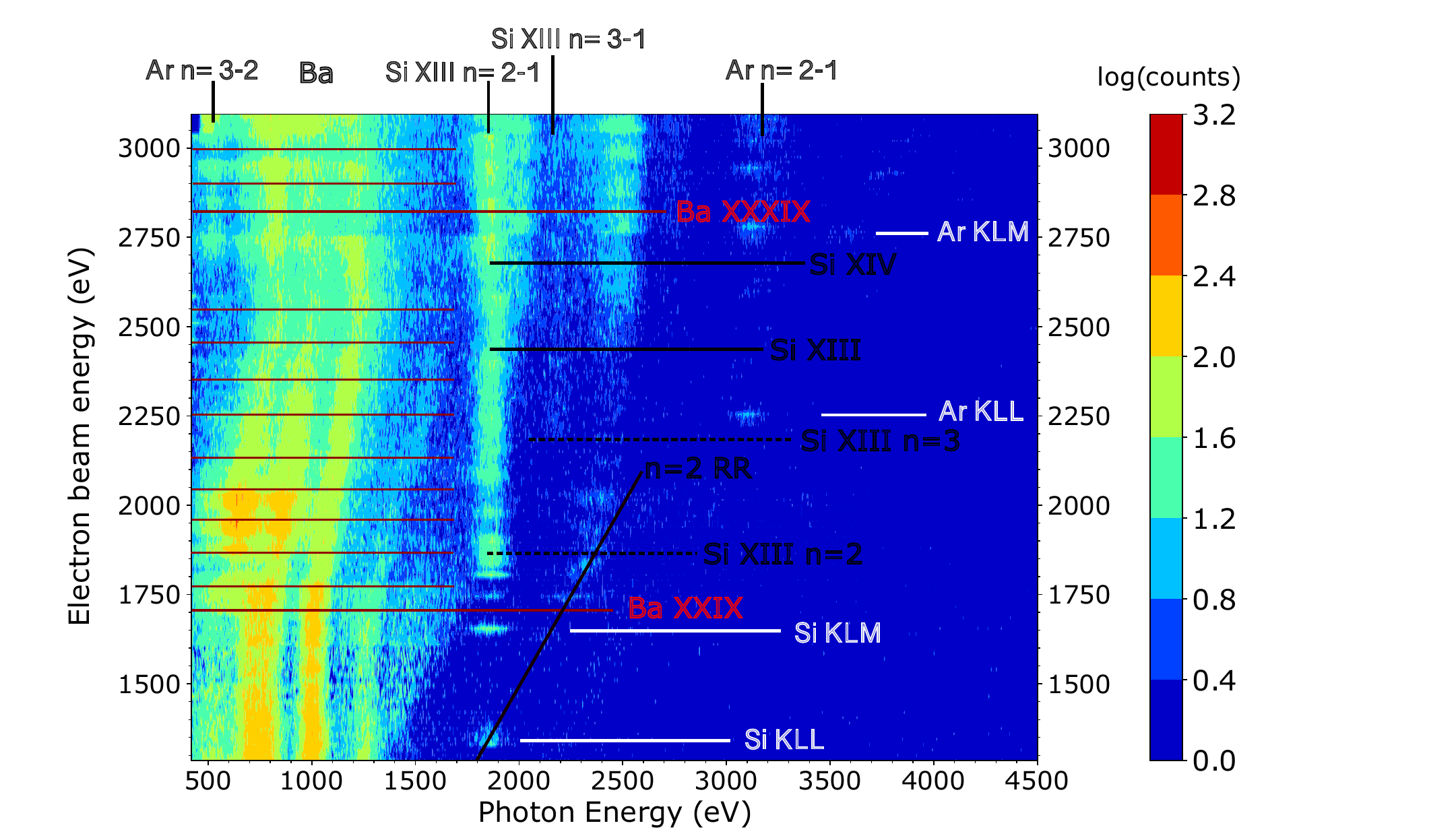}
    \caption{Contour plot showing background spectra taken over a range of electron beam energies. Data were corrected for any differences in collection time or electron beam current. Si and Ar DR features are labeled with white horizontal lines at the corresponding beam energy. Horizontal black dotted lines show the $n = 2-1$ and $n = 3-1$ transition energies of Si XIII (He-like). Solid black, horizontal lines show the ionization potential of He-like (XIII) and H-like Si (XIV). Vertical lines at the top indicate the photon energy of Ar and Si transitions. The ionization potential for Ni-like Ba (XXIX, 1695 eV) to S-like Ba (XLI, 2994 eV) are shown as red horizontal lines across Ba features between 500 -1700 eV photon energy. Diagonal solid black line traces the radiative recombination emission into the $n = 2$ shell of He-like Si. }
    \label{fig:bacontour}
\end{figure*}


\subsection{\label{sec:results} Discussion and Results of Emitted Spectra}

 The spectra of EBIT background elements are shown in Fig. \ref{fig:bacontour}. The y-axis is the non-space-charge corrected electron beam energy, although at the currents, energies, and magnetic field used, the space charge correction is minimal \citep{Porto_2000}. In Fig. \ref{fig:bacontour}, the vertical emission lines result from electron impact excitation and faint diagonal lines are formed by radiative recombination.
 
 Dielectronic recombination is a three-step resonant process by which a free continuum electron is captured by an ion while simultaneously exciting a bound electron, placing the recombined ion in a doubly-excited state. The ion can stabilize by either autoionization or through radiative decay of the excited electrons. The capture process only occurs when the sum of the energy of the free electron and the binding energy of the recombined ion equals the excitation energy of the bound electron. When creating this process with a mono-energetic electron beam, signature bright emission spots corresponding to specific electron and and photon energies, as seen in Fig. \ref{fig:bacontour}, are created below the direct excitation (DE) threshold. 
 
 The low 500 eV - 1500 eV photon energy vertical and wavy lines are created from Ba ions. Ni-like Ba (Ba XXIX), having a closed 3d shell, is produced at electron beam energies above 977 eV \citep{NIST_ASD}, while neighboring Co-like Ba is not created until beam electrons have an energy of at least 1695 eV. Therefore, vertical Ba lines between 1285 eV and 1700 eV (electron beam energy) are primarily due to 4p - 3d and 4f - 3d transitions in Ni-like Ba. 
 
Above 1700 eV emission originates from 4p - 3d and 4f - 3d transitions in Co-like Ba. With an ionization energy of the remaining 3d ions being roughly 100 eV apart, the photon energies of the 4p - 3d and 4f - 3d transitions slowly shift to higher energies with the higher charge states (see Fig. \ref{fig:bacontour}). Finally, at an electron energy of 2547 eV, K-like Ba is ionized and Ar-like Ba (Ba XXXIX), having a closed 3p ground shell, is created. This charge state dominates above a beam energy of about 2600 eV.
 \begin{figure*}[!tbp]
  \centering
  \includegraphics[width=0.45\textwidth]{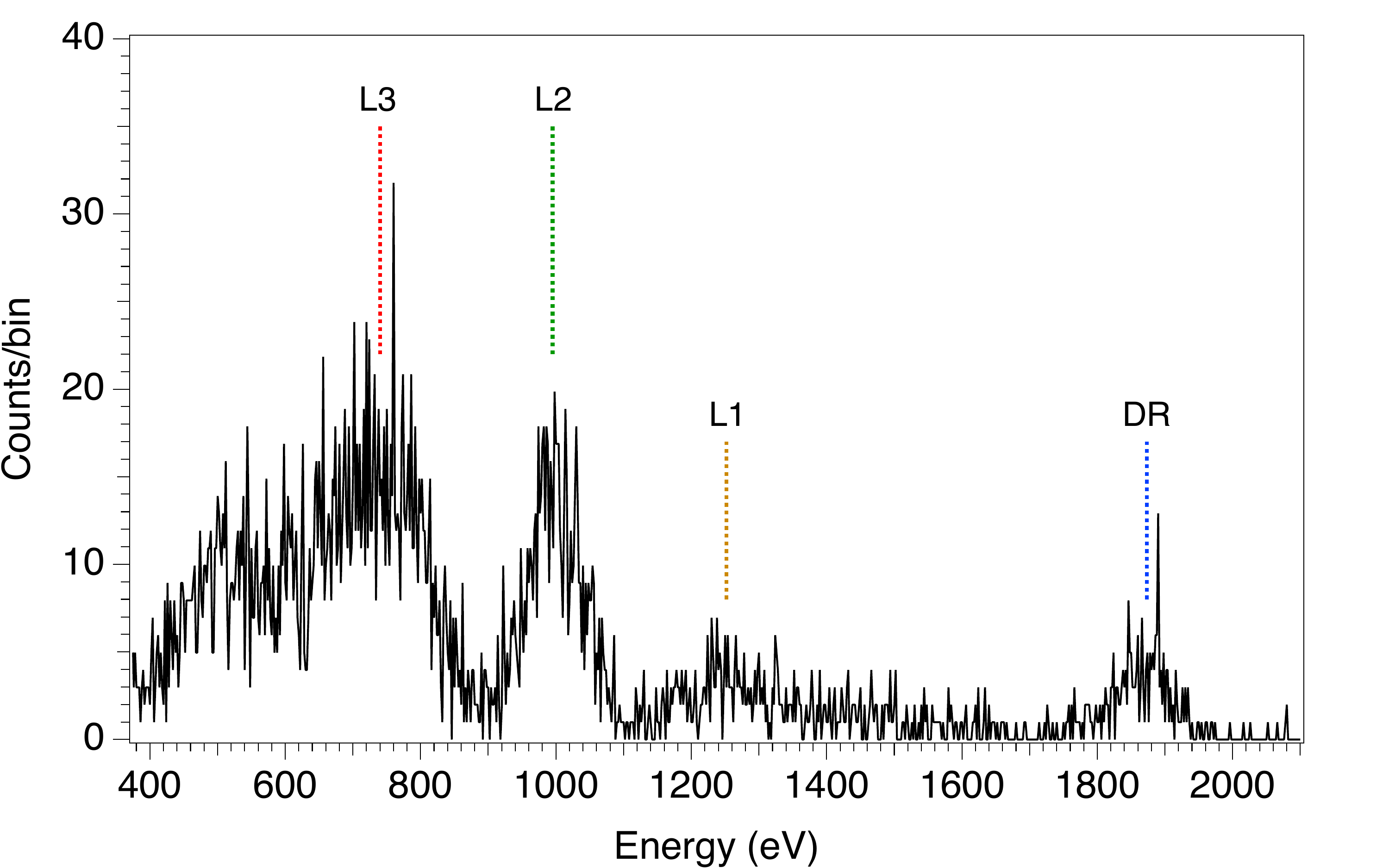}
  \hfill
  \includegraphics[width=0.45\textwidth]{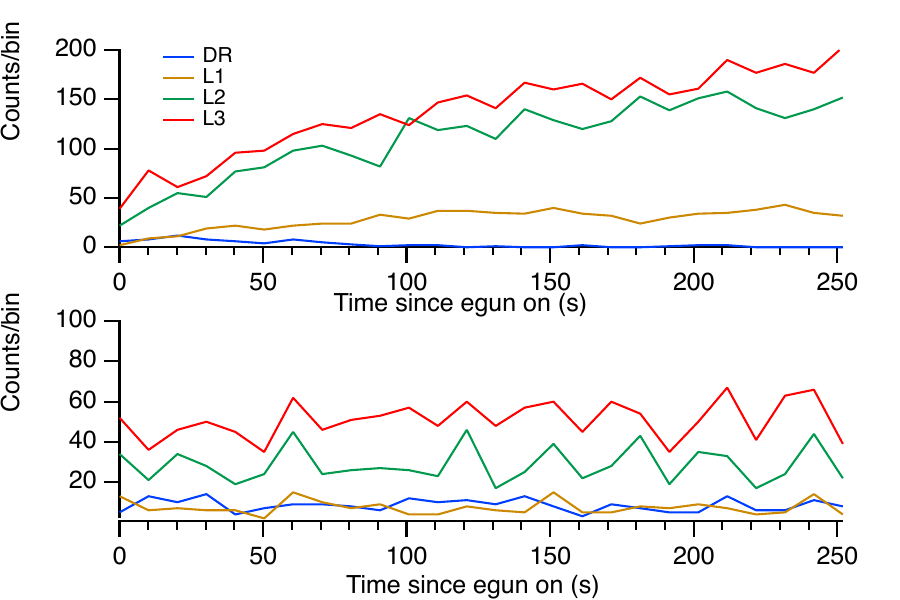}
  \caption{(Left) Spectrum taken at an electron beam energy of 1650 eV, near the Si KLM maximum. Labels indicate the Si DR (DR) at 1873 eV, and three Ba lines at 1252 eV (L1) , 995 eV (L2), and 740 eV (L3)(Right) Sum of the counts in 10 s bins from the Si KLM DR feature and three Ba lines shown in the left figure
  Top: no dumping of the trap. Bottom: dumping the trap every 60s.}
  \label{fig:1650spec}
\end{figure*}
He-like Si (Si XIII) is created in this electron beam energy range above 523 eV \citep{NIST_ASD}, and the $n = 2 - 1$ transition energy is around 1865 eV (shown as a dotted horizontal line in Fig. \ref{fig:bacontour}). Below this threshold beam energy, strong KLL and KLM (in inverse Auger process notation \citep{Knapp_1993}) DR features appear, while above the threshold, emission comes from direct excitation. The peak intensities of the Si KLL and KLM (n=2-1 transitions) features occur at beam energies of 1330 eV and 1660 eV, respectively. Si emission above the He-like Si ionization threshold comes from both H-like and He-like charge states.

In addition to Si and Ba emission, features from lingering Ar were also observed. Ar KLL and KLM DR satellites from different charge states appear at electron beam energies below their respective DE thresholds ($\sim$3100 eV). The corresponding emitted photon energies are above 3100 eV  for $n \geq 2-1$ transitions and around 520 eV for $n = 3-2$ transitions. Both the Ar and Si DR features appear at beam energies consistent with previously reported resonance energies, further emphasizing the negligible space charge effect for this work \citep{Gall_2019,Baumann_2014}.

\subsection{\label{sec:dump} Optimizing the Trapping Time}
During measurements the drift tube electrodes were set to 350 V, 0 V, and 150 V for DT 1, 2, and 3, respectively. At the end of the trapping period, the voltage of the middle drift tube was raised to 250 V for 5 ms  to empty or dump the ions from the trap and increase the signal from DR features.

\begin{figure*}[!tbp]
  \centering
  \includegraphics[width=0.45\textwidth]{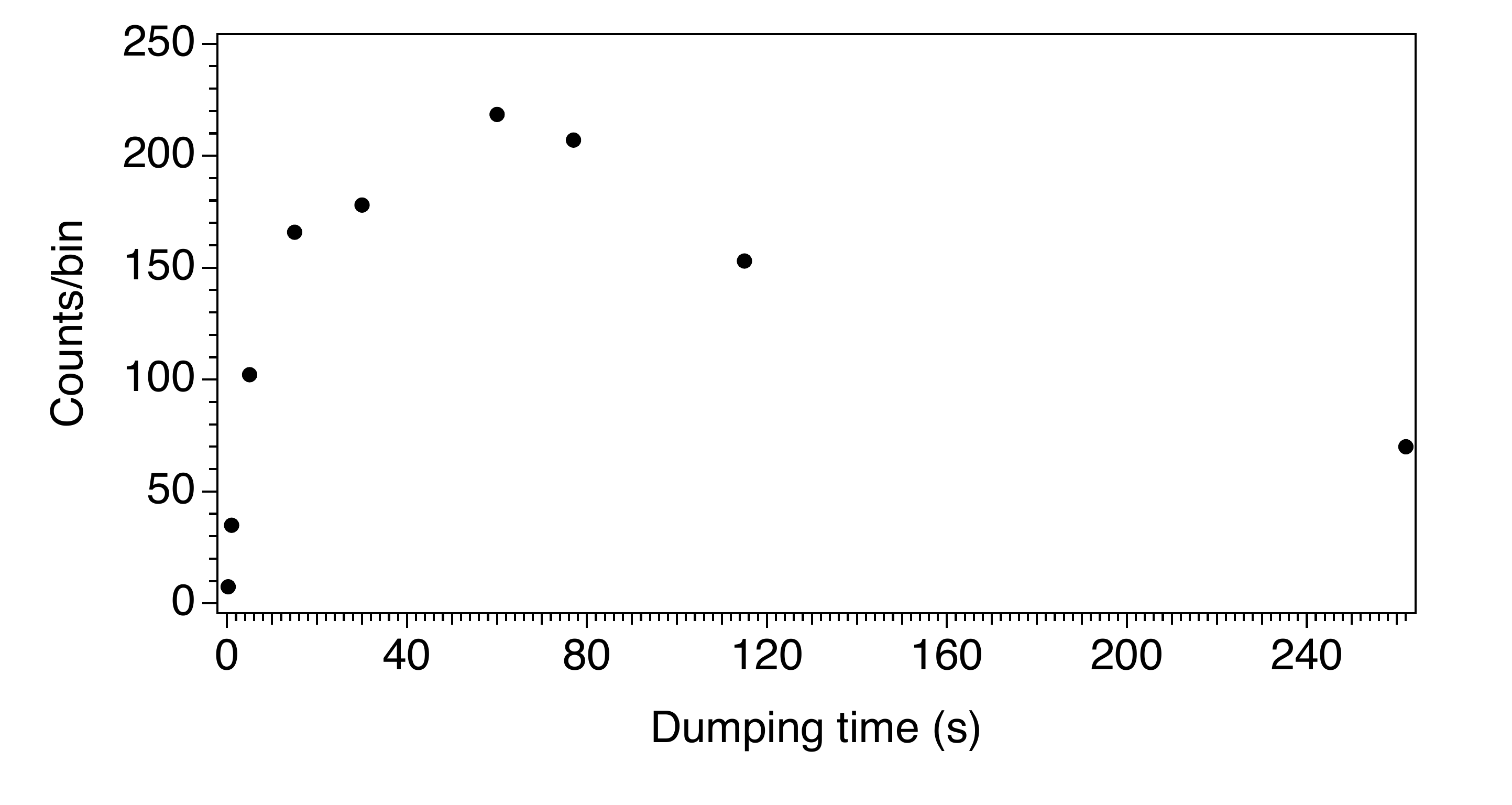}
  \hfill
 {\includegraphics[width=0.45\textwidth]{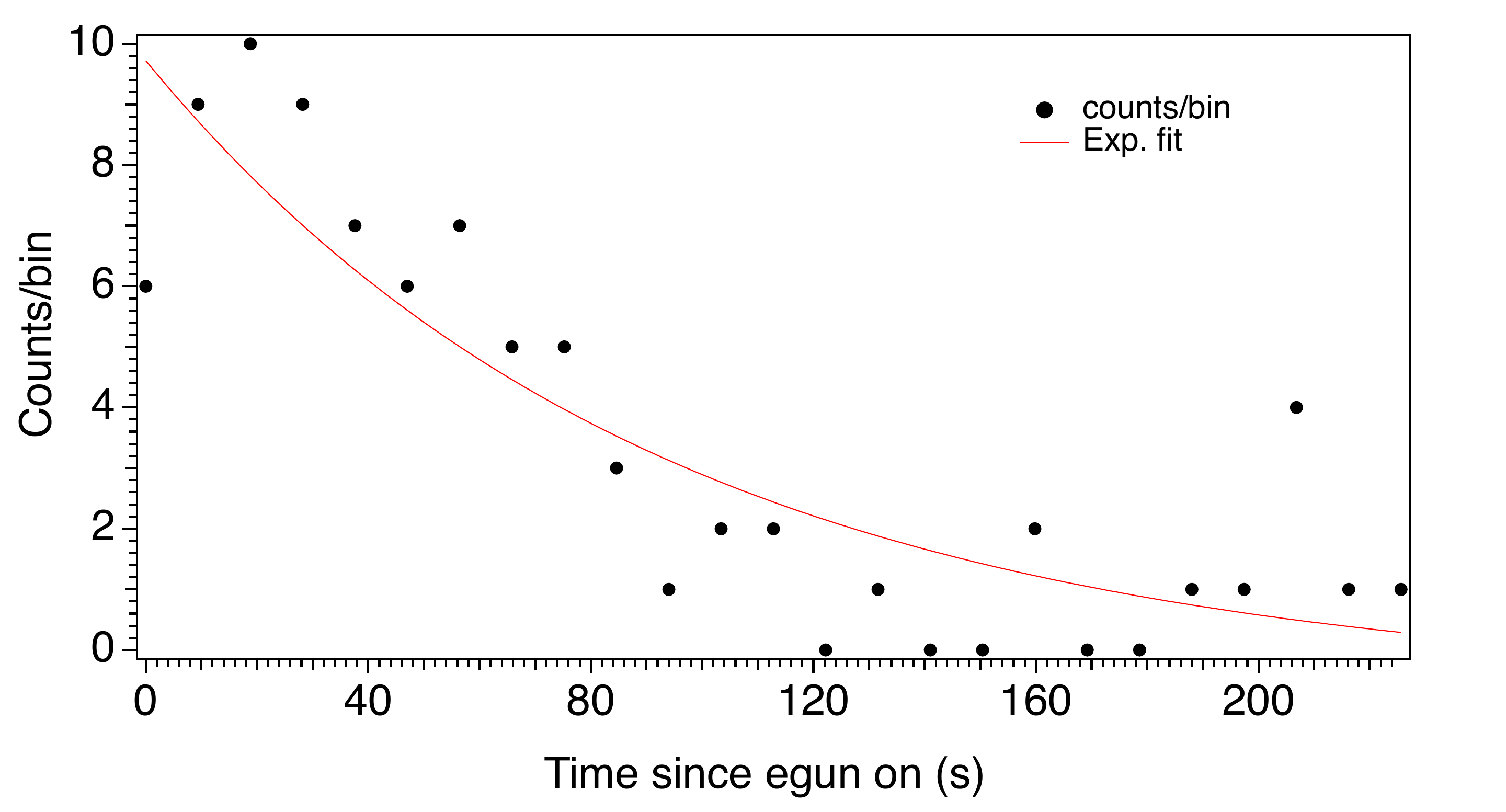}}
  \caption{(Left) Each data point represents the Si DR peak counts summed over a single measurement with different dumping times, ranging from every 0.5 s to no dump (260s measurement). Data were corrected for any differences in total collection time. (Right) Single 225 s measurement, with no dumping of the trap, broken up into 10 s time bins. Counts in the strong Si DR peak were summed for each 10 s bin, showing that the Si DR peak counts fall off exponentially with time.}
  \label{fig:decay}
\end{figure*}

Figure \ref{fig:1650spec} (Left) shows the spectrum taken at an electron beam energy of 1650 eV, where the Si DR peak is near its maximum intensity. Three Ba lines (L1, L2, and L3) are labeled along with the Si DR peak (labeled DR). Counts in the Si DR peak were compared for measurements taken with various dump-cycle times. Based on the results shown in Fig. \ref{fig:decay} (Left), a dump time of 60 s was used, as this produced the maximum signal.  
 
During measurements each X-ray photon is time-stamped. Plotting the counts in the three Ba lines (L1, L2, L3) with the Si DR feature as a function of time (10 s bins) for the case of no dumping and with 60 s dumping shows that the count rate of the four lines is stable when the trap is dumped every 60 s (Fig. \ref{fig:1650spec} (Right)). However with no dumping, the Si DR counts exponentially decrease with time (shown more clearly in Fig. \ref{fig:decay} (Right)), the L1 line exponentially increases and stabilizes, and L2 and L3 rapidly increase at a similar rate.

While all of the trapped ions are heated through collisions, the lowest charged ions escape from the trap first (escape temperature is proportional to the charge \citep{Levine_1988}). This is the mechanism used in evaporative cooling, where light elements are injected into the trap to interact with highly charged ions and carry away energy as high energy particles leave the trap \citep{Currell_2005} Therefore, Fig. \ref{fig:1650spec} can be explained by the light Si ions quickly disappearing from the trap while highly charged Ba ions continue to accumulate. This also demonstrates that the charge state balance is not in equilibrium when the trap is not dumped. 

\section{\label{sec:plans} Conclusion and Outlook}
The SAO EBIT facility is fully operational again after downtime following a silicone oil contamination event and thorough decontamination of the instrument. Background elements from the electron gun, residual gas, and elements from trace amounts of Si oil residue have been measured and the evolution of trapped ions characterized. Updates to the EBIT thermal system and the addition of new injection methods 
improve the system stability and the range of possible scientific studies. 

X-ray slit and optical lens setup have been developed to measure the spatial distribution of the electron beam and the ion cloud. From the measured FWHM of the beam and the ion cloud, the effective electron density has been determined and found to be about an order of magnitude lower than determined assuming a uniform cylindrical beam, in agreement with observations from other EBIT devices. In the future this setup can be used to carry out calibration of density sensitive line ratios important in astrophysics. Furthermore, from the estimates of the size of the electron beam and effective density,  estimated count rates can inform future studies and detector system designs, such as slit-less spectrometers where the electron beam provides a slit like source.

To support the needs of XRISM, a high-resolution X-ray microcalorimeter, built originally by SAO for the NIST EBIT \citep{Silver_2005}, was recently brought back to SAO. The detector has been fully tested with the X-ray calibration source and integration and initial measurements with the EBIT are upcoming. 

In an effort to better understand measured features and the population dynamics of highly charged ion plasma, we are currently developing a collisional radiative (CR) model. The non-Maxwellian CR model will allow us to create synthetic EBIT spectra to aid in line identifications and to test the model's underlying atomic data (which can be easily changed). A detailed description of the model and comparisons with EBIT measurements will be provided in a separate publication.  

While there is no extraction system currently installed, the horizontally-oriented SAO EBIT is ideally suited for ion extraction. Long term plans include the design of such a system that will provide a direct probe of the ion charge state distribution, allow for the creation of ion beams of select charge states for charge exchange,  surface interaction studies, or re-trapping of ions.


\bibliography{refs}{}
\bibliographystyle{aasjournal.bst}

\end{document}